\documentclass[superscriptaddress,preprint,amsmath,amssymb]{revtex4}

\usepackage{graphicx}
\usepackage{braket}

\begin{document}
\title{Beneficial impact of tunneling in nano-structured intermediate-band solar cell}
\author{Nicolas Cavassilas}
\affiliation{Aix Marseille Universit{\'e}, CNRS, Universit{\'e} de Toulon, 
IM2NP UMR 7334, 13397, Marseille, France}
\affiliation{NextPV, LIA, CNRS-RCAST/U. Tokyo-U. Bordeaux, Tokyo 153-8904, Japan}
\author{Daniel Suchet}
\author{Amaury Delamarre}
\affiliation{Research Center for Advanced Science and Technology, The University of Tokyo, Tokyo 153-8904, Japan}
\affiliation{NextPV, LIA, CNRS-RCAST/U. Tokyo-U. Bordeaux, Tokyo 153-8904, Japan}
\author{Fabienne Michelini}
\affiliation{Aix Marseille Universit{\'e}, CNRS, Universit{\'e} de Toulon, 
IM2NP UMR 7334, 13397, Marseille, France}
\author{Marc Bescond}
\affiliation{LIMMS, CNRS-Institute of Industrial Science, UMI
2820, University of Tokyo, 153-8505 Tokyo, Japan
}
\author{Yoshitaka Okada}
\author{Masakazu Sugiyama}
\affiliation{Research Center for Advanced Science and Technology, The University of Tokyo, Tokyo 153-8904, Japan}
\affiliation{NextPV, LIA, CNRS-RCAST/U. Tokyo-U. Bordeaux, Tokyo 153-8904, Japan}
\author{Jean-Francois Guillemoles}
\affiliation{NextPV, LIA, CNRS-RCAST/U. Tokyo-U. Bordeaux, Tokyo 153-8904, Japan}
\affiliation{IRDEP, UMR 7174 CNRS EDF Chimie ParisTech, EDF R$\&$D, Chatou, France}

\begin{abstract}
Using the non equilibrium Green functions formalism we propose a study of the electronic excitation and collection in nano-structured intermediate-band solar cell. We demonstrate that a thin tunnel barrier between the nano-objects and the host material is beneficial for both current and voltage. For the current, the confinement generated by such a thin barrier favors the intersubband optical coupling in the nano-objects and then improves the excitation-collection trade-off. We also show that a broaden density-of-state in the nano-objects increases the radiative recombination and then degrades the voltage. Using a detailed balance model we propose a \textit{broadening factor} for this $V_{oc}$ degradation which decreases when a tunnel barrier enhances the life-time in the nano-objects.
\end{abstract}

\pacs{Valid PACS appear here} 
\maketitle
\section{Introduction}
By introducing an intermediate band (IB) in a wide bandgap solar cell, the aim is to increase the short-circuit current $I_{sc}$ without reducing the open-circuit voltage $V_{oc}$ \cite{luque_understanding_2012,luque_increasing_1997}. The current increases as the IB acts as a built-up converter to generate an electron-hole pair with two low-energy photons. To limit detrimental effect voltage drop, the IB must be electrically isolated from the contacts of the cell so that the recombination is still controlled by the wide bandgap. This theoretically allows to exceed the Shockley-Queisser (SQ) limit \cite{shockley_detailed_1961} with a $V_{oc}$ controlled by the wide bandgap while low-energy photons generate electron-hole pairs.

Yet, practically, this concept has not yet resulted in a cell exceeding the SQ limit \cite{sogabe_intermediate-band_2014,okada_intermediate_2015}. Worse, IB solar cell (IBSC), generally based on nano-objects, offers characteristics that are often lower than that of the same cells without IB \cite{pusch_limiting_2016}. The small increase of current is not enough to catch up with the degradation of the $V_{oc}$. It is difficult to obtain a splitting between the Fermi levels of the IB and the contact (usually the n-type contact) because the electrons easily relax from the conduction band of the wide bandgap material to the IB. In other words, the IB is not well isolated from contacts and $V_{oc}$ is controlled by recombination across a narrow bandgap.

Furthermore the increase of current remains low as it is difficult to have simultaneously an efficient photon absorption and a fast collection of the excited electrons. Indeed, if the IBSC is based on nano-objects, the electrons in IB are confined in localized states. The optical transition between the IB and the conduction band is then efficient if the excited state is also localized. This improves the wave-function overlap and the corresponding intraband (intersubband) transition \cite{bastard_wave_1988}. However, if the excited electron is strongly confined in the nano-object, it cannot reach in the conduction band to be collected. It will finally relax by emitting photon and/or phonons. As a result the choice of the coupling between the excited state of the nano-objects and the conduction band of the wide bandgap is crucial to have a good excitation-collection trade-off. In case of bound-to-bound absorption with a thick tunnel barrier the absorption is high, but the electrons collection is limited. In a bound-to-continuum absorption without tunnel barrier the collection is fast but the absorption is low. In this letter we propose a comprehensive analysis of this problem. We will see that a thin tunneling barrier between the nano-object and the wide bandgap material offers a good excitation-collection trade-off. This permits to optimize the current. More unexpected, we will show that such a coupling also modifies the photon emission and then the $V_{oc}$. 

\section{Model and system}

This theoretical study is based on a quantum electronic transport model in the non equilibrium Green functions (NEGF) framework. This model allows to consider the quantum effects such as the confinement, the tunneling, the broadening due to the coupling and the scattering with the photons and the phonons. Nevertheless, except when this will be explicitly stated, we will not consider scattering with phonons. Such a radiative limit permits to facilitate the understanding of the results. The model used is described in Ref. \cite{cavassilas_modeling_2013} but as suggested in Ref. \cite{cavassilas_local_2016} we now consider the non-local character of the electron-photon interaction which guarantees compliance with the selection rules.

The modeled system is schematically shown Fig. 1. We consider a quantum well (QW) between two contacts. The left contact can only inject and collect electrons over an interval of 0.2 eV around the left Fermi level $\mu_L$. The latter is located at the maximum of the density-of-states (DOS) of the ground state of the well. While this ground state mimics the IB, the corresponding Fermi level mimics the absorption between the valence band (no assumed in this work) and the IB. The electrons on this ground state can interact with photons and reach an excited state. From there, they either relax (and then do not participate in the current) or reach the right contact. This contact, located at an energy $E_c$ above $\mu_L$, represents the conduction band-edge of the wide bandgap material. By reaching this contact the electrons generate a current that one chooses positive. The right contact can also inject electrons that can reach the QW, relax by emitting a photon and then reach the left contact. This behavior, which corresponds to the relaxation from conduction band to IB, generates a negative current which is all the more important as the right Fermi level $\mu_R$ is high. In an IBSC the aim is to have a positive current balance (more excitations than relaxations) while having a higher Fermi level in conduction band ($\mu_R$) than in IB ($\mu_L$). Our system therefore behaves like an intraband solar cell under bias $qV=\mu_R-\mu_L$ and characterized by a current $I_{sc}$ and a voltage $V_{oc}$.

\begin{figure}[h!]
\centering
\includegraphics[width=0.45\textwidth]{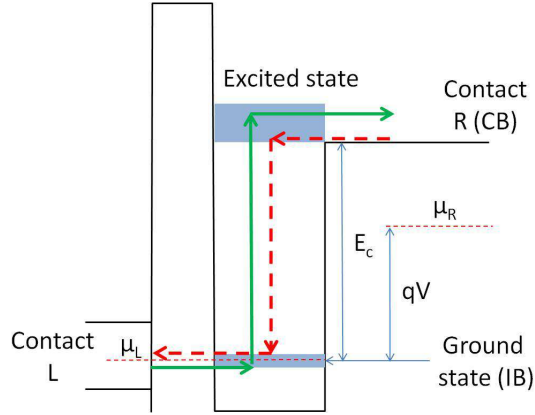}
\caption{Schematic representation of the modeled system. The full-arrows shows he radiative excitation and the collection while the dash-arrows represent the injection and the emission. We also shows the two Fermi levels $\mu_R$ and $\mu_L$ respectivelly correspondind to the IB and the n-type contact. The grey surfaces represent the states.} \label{F1}
\end{figure}

\begin{figure}[h!]
\centering
\includegraphics[width=0.45\textwidth]{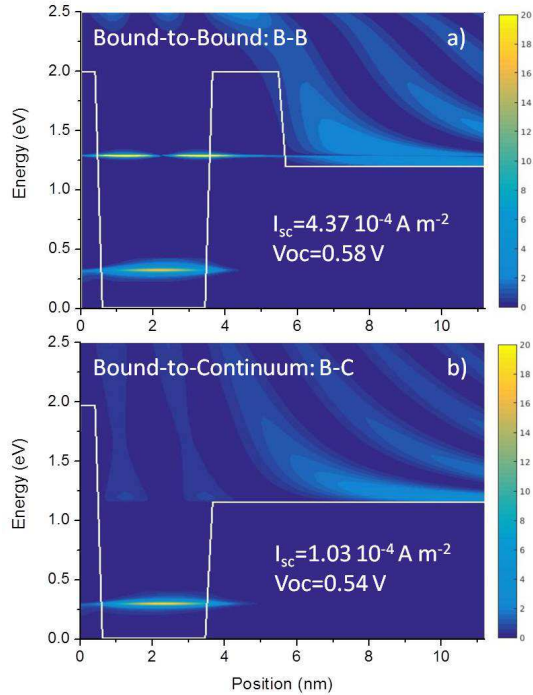}
\caption{The band-diagram (white line) and the local-DOS of electrons in a) the Bound-to-Bound system and in b) the Bound-to-Continuum system.} \label{F2}
\end{figure}

In the following we compare a bound-to-bound system in which a tunnel barrier is assumed between the QW and the right contact to a bound-to-continuum system in which we do not assume any barrier. While for an interband transition the definition of the energy gap seems obvious, this is not so trivial for an intraband-one. We defined the gap as the energy of the maximum absorption considering a black-body at 6000 K for the incident photon flux. For all systems assumed in this work, the gap $E_g$ is the same. We chose $E_g=1$ eV which is large compared to what is generally considered in IBSC. This choice is based on a study conducted elsewhere \cite{amaury} which shows that with a ratchet mechanism \cite{pusch_limiting_2016} of 0.7 eV, such a value is optimal in the case where the absorption is rather narrow as expected in intraband system. The other parameter is the effective mass which equals 0.665 (in free electron mass). Finally  we chose to treat a QW since our one-dimensional model is well adapted to such a system. However, the physical behaviors such as confinement and absorption are very close in case of quantum dots. Moreover, if it is straigthforward to add a barrier between a QW and a contact, it is also feasible in case of quantum dots assuming a core-shell architecture \cite{kim_engineering_2005} or dots embedded in specific material \citep{sato_extremely_2012}. If in the present study we assume a QW, the conclusions can thus be easily extended to the quantum dots.

\section{Results}
\subsection{Current and voltage with the quantum model}
The band diagrams and the local-DOS of the two systems are represented in Fig. 2. In both the well thickness is 3.2 nm. In one case, the excited state is a localized state separated from the right contact by a 2 nm-thick tunnel barrier (Fig 2a). In this system, called bound-to-bound (B-B), $E_c$ equals 0.88 eV. In the other system the electrons are excited in an energy continuum directly connected to the right contact (Fig 2b). In the latter case, called bound-to-continuum (B-C), in order to have $E_g=1$ eV, $E_c$ is reduced to 0.855 eV. On Fig. 2 the LDOS clearly shows that in B-B the electrons are excited in a localized and narrow quantum state. In B-C, despite the absence of barrier, we do not obtain a \textit{true} continuum but rather a strongly broadened state. Indeed, even without barrier, the QW involves quantum reflections and then interferences of the electronic wave-function.

\begin{figure}[h!]
\centering
\includegraphics[width=0.45\textwidth]{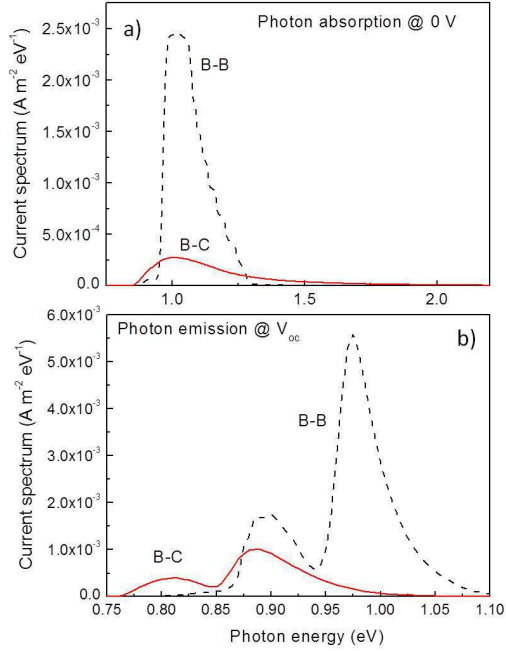}
\caption{Current-density \emph{versus} the photon energy generated in each structure B-B and B-C by a) photon absorption and b) photon emission.} \label{F3}
\end{figure}

Regarding the results Fig. 2, $I_{sc}$ and $V_{oc}$ are both higher in B-B. In order to better understand this results, Fig. 3 shows the electronic current \textit{versus} the photon energy for the two systems. The absorption spectra corresponds to the positive electronic current while the emission is the negative component. The absorption in B-B is narrower since, as shown in Fig. 2, the excited state is thinner. At same time this absorption is much higher than its the B-C counterpart. This can be explained by a better trade-off of the electronic collection and of the radiative excitation due to a larger wave-functions overlap in the bound-to-bound configuration. This explains that $I_{sc}$ is larger in B-B despite the tunneling.

From the emission point of view, we obtain a lower bandgap in B-C. This behavior, which explains the $V_{oc}$ degradation in B-C, may have several origins. In B-C, photons with lower energy are emitted since $E_c$ is lower. More relevant, due to tunneling in B-B, the emission of photon with low energy (between 0.78 and 0.85 eV) is reduced. But at that point, the origin of the lower emission bandgap  in B-C is not yet clear.

It is possible to improve B-C, for example, by reducing the thickness of the well (2.8 nm). This permits, as shown Fig. 4, to have a quasi-bound state narrower than the strongly broaden state of the B-C. In this new device, called B-Q, by adjusting $E_c$ at 0.88 eV in order to have $E_g=1$ eV, the current is largely higher than in the original B-C device, while $V_{oc}$ is quite similar. Such a current increase is due to a better excitation-collection trade-off. However the current remains lower than in B-B. Moreover, the fact that in B-C and B-Q $V_{oc}$ is close and lower than in B-B, confirms that $V_{oc}$ is not simply controlled by $E_c$. More generally such results suggest that a narrow excited state is better for both $I_{sc}$ and $V_{oc}$. In the following, we propose to use a simpler model in order to verify and to explain this unintuitive feature.

\begin{figure}[h!]
\centering
\includegraphics[width=0.45\textwidth]{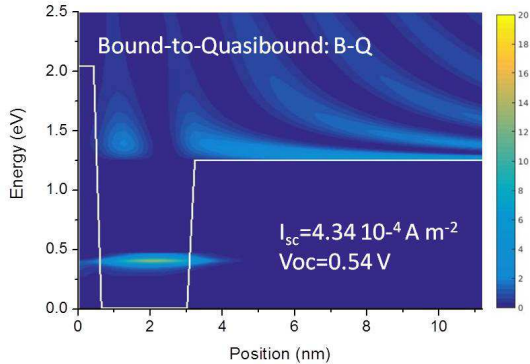}
\caption{The band-diagram (white line) and the local DOS of electrons in the Bound-to-Quasibound system.} \label{F4}
\end{figure}

\subsection{\textit{Broadening factor} in the detailed balance}
We assume a detailed balance model where absorption and emission are calculated assuming a black-body distribution for the photons and a Lorentzian shape for the electronic DOS. Indeed, if we assume an homogeneous broadening due to contact (or others), the DOS spectral shape of states in the QW is given by a Lorentzian distribution \cite{bastard_wave_1988}. If the tunneling between the excited state and the contact is low (large life-time $\tau$) the width $\Gamma$ of the Lorentzian is low following $\Gamma=\hbar/\tau$. 
We then apply this to the DOS of the exited state while, in order to simplify the model, the ground state is assumed as a Dirac function. We finally calculate the current, the emission and the voltage as:

\begin{eqnarray}
&I_{sc}&=\int_{E_c}^{\infty}A_b(E)dE\nonumber\\
&&=fM^2\int_{E_c}^{\infty}\rho(E,T_{sun},\Delta\mu=0)L(E,\Gamma,E_g)dE
\end{eqnarray}

\begin{eqnarray}
&E_m&=\int_{E_c}^{\infty}E_m(E)dE\nonumber\\
&&=\pi M^2\int_{E_c}^{\infty}\rho(E,T_{cell},\Delta\mu=qV)L(E,\Gamma,E_g)dE
\end{eqnarray}
\begin{equation}
V_{oc}= k_BT_{cell}\log\left(\frac{I_{sc}}{E_m(\Delta\mu=0)}\right)
\end{equation}
 
with the black-body distribution
\begin{equation}
\rho(E,T,\Delta\mu)=\frac{2}{h^3c^2}\frac{E^2}{\exp(\frac{E-\Delta\mu}{k_BT})-1},
\end{equation}

and the electronic joined DOS
 \begin{equation}
L(E,\Gamma,E_g)=\frac{\frac{1}{\pi\Gamma}}{1+\left(\frac{E-E_g}{\Gamma}\right)^2},
\end{equation}
where $\Gamma$ is the width, $E_g$ the energy of the maximum of the Lorentzian, $qV=\mu_R-\mu_L$ the bias, $k_B$ the Boltzmann constant, and $f=C\times 6.79\times 10^{-5}$ with $C=1$ the concentration factor. Note that for the integral of $A_b$ and $E_m$ the minimum of energy is given by $E_c$ the band-edge of the right contact.  Finally $T_{sun}$ and $T_{cell}$ are respectively the temperature of the sun (6000 K) and of the cell (300 K). In this model the excitation-collection trade-off is not considered since $M^2$, which represents the optical coupling, is taken as a constant and $I_{sc}$ is directly given by the photon absorption in Eq. (1). This model is schematically described in Fig. 5.

\begin{figure}[h!]
\centering
\includegraphics[width=0.45\textwidth]{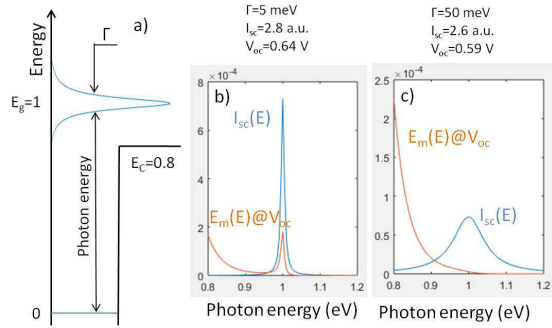}
\caption{a) The schematically described detailed balance model and the corresponding $I_{sc}$ and $E_m$ spectra for b) $\Gamma=5$ meV and for c) $\Gamma=50$ meV. }
\label{F5}
\end{figure}

Figure 5 also shows the spectra $A_b(E)$ and $E_m(E)$ for two values of $\Gamma$ (5 and 50 meV) with $E_g=1$ eV and $E_c=0.8$ eV.  As obtained with the NEGF model, and although this detailed balance model does not take into account the excitation-collection trade-off,  we obtain both $I_{sc}$ and $V_{oc}$ larger with a thinner state. As shown Fig. 5, $I_{sc}$ and $V_{oc}$ are even more degraded than the DOS is high in the QW at the band-edge of the contact ($L(E_c,\Gamma,E_g)$). Indeed, a large $L(E_c,\Gamma,E_g)$ means that a large proportion of excited electrons cannot be collected since their energy is lower that the band-edge of the contact. This degrades the current. A large $L(E_c,\Gamma,E_g)$ also means a large electronic injection from the contact at low energy where the emission is very efficient. As already observed Fig. 3, this reduces the bandgap energy of the photon emission and then degrades the voltage. Such a behavior, as well as for example the reduction of the bandgap by tunneling in ultra thin cells \cite{aeberhard_microscopic_2017}, shows that the electronic transport modifies the optical properties of nanoscale devices. With the detailed balance model we finally obtain this approximate expression for $V_{oc}$: 
 \begin{eqnarray}
&V_{oc}\simeq&\left(1-\frac{T_{cell}}{T_{sun}}\right)E_g-\frac{k_BT_{cell}}{q} \log\left(\frac{\pi}{f}\right)\nonumber\\&&-\frac{k_BT_{cell}}{q} \log\left(1+A\Gamma\right)
\end{eqnarray}
with
\begin{eqnarray}
&A=&\frac{k_{B}T_{cell}\,\left(E_{c}^{2}+2k_{B}T_{cell}\,E_{c}+2\left(k_{B}T_{cell}\right)^{2}\right)}{\pi E_{g}^{2}\left(E_{g}-E_{c}\right)^{2}}\nonumber\\&&\exp\left(\frac{E_{g}-E_{c}}{k_{B}T_{cell}}\right).
\end{eqnarray}
The two first terms in the expression of $V_{oc}$ are the well-known Carnot and Boltzmann factors \cite{markvart_solar_2008,hirst_fundamental_2011} while the third one is relative to an original broadening behavior that we call the \textit{broadening factor}. The origin of this degradation, as already observed in bulk materials \cite{yao_quantifying_2015,rau_efficiency_2017}, is the mismatch between the absorption and the emission bandgaps. This factor is generally no considered in detailed balanced \cite{markvart_solar_2008} since the absorption is assumed sharp at the bandgap. In intraband system this broadening behavior degrades both $I_{sc}$ and $V_{oc}$ when the excited state is too thick like in the bound-to-continuum configuration. 

Phonon emission is expected to favor the band-edge photon emission and then to degrade $V_{oc}$. With our NEGF model we show that such scattering with polar optical phonon is stronger in the continuum than in the QW (and is expected to be even weaker in quantum dot). The consequence is that scattering degrades $V_{oc}$ in B-C (-90 mV) while this degradation is not significant in B-B. This result suggests that scattering with phonon increases the \textit{broadening factor}. However this point deserves further investigations which cannot be conducted in the present work due to huge numerical burden.

\begin{figure}[h!]
\centering
\includegraphics[width=0.45\textwidth]{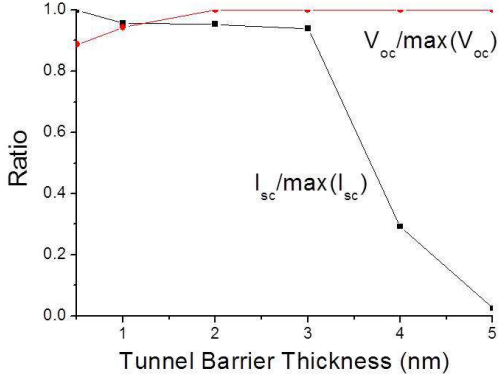}
\caption{$I_{sc}/\mathrm{max}(I_{sc})$ and $V_{oc}/\mathrm{max}(V_{oc})$ \textit{versus} the thickness of the tunnel barrier $t_B$.Assuming a thick barrier excited states are very thin and energy mesh has to be also very thin. From a computational point of view we cannot conduct the calculation for tunnel thicker than 3 nm assuming the transverse dispersion. This calculation has then been conducted only for a transverse wave vector equals to zero.} \label{F6}
\end{figure}

\subsection{Optimization of the tunnel barrier}

The \textit{broadening factor} may suggest that in the B-B system better characteristics are expected if we assume a thick tunnel barrier between the nano-object and the contact. Indeed, due to strong tunnel reflection, life-time increases and then broadening of the excited state decreases. Fig. 6 shows the evolution of both $I_{sc}$ and $V_{oc}$ \textit{versus} the tunnel barrier thickness $t_B$ in the B-B system (remember that in Fig. 2a, $t_B$=2 nm). For $t_B$ lower than 2 nm, the reduction of $V_{oc}$ is due to the \textit{broadening factor}. At the same time the larger current with the ultra-thin barrier suggests another behavior like a reduction of the tunnel reflection improving the excitation-collection trade-off. This is confirmed for the thick barriers where $I_{sc}$ and then the collection is strongly degraded by such a reflection. With a barrier thicker than 3 nm the excitation-collection trade-off is strongly degraded. In the same time, $V_{oc}$ no more increases for barrier thicker than 2 nm since the excited state is thin enough ($\le 5$ meV) to cancel the \textit{broadening factor}. For an optimal tunnel barrier both \textit{broadening factor} and excitation-collection trade-off should be considered. With the parameters considered in this work the trade-off is optimal for a 2 nm-thick tunnel barrier. In case of an other material with an effective mass $m^*$ (in free electron mass) and a rectangular tunnel barrier offset $\Delta$ (in eV), this tunneling coupling should be conserved if the thickness of the barrier (in nm) is given by $t_B=\frac{0.43}{\sqrt{m^*\Delta}}$.

\section{Conclusion}

We have shown that a \textit{broadening factor} degrades the voltage in intraband system by improving the photon emission. To avoid this degradation a bound-to-bound system in which the electrons are collected by tunneling is well adapted. Moreover, such a thin tunnel barrier improves the excitation-collection trade-off due to a higher wave-function overlap. On the other hand a too thick tunnel barrier degrades this excitation-collection trade-off due to strong tunnel reflection. It is then necessary to choose a tunnel barrier thick enough to cancel the \textit{broadening factor} and to improves the wave function overlap, but, thin enough to preserve the excitation-collection trade-off. We finally propose an equation allowing to calculate the thickness of the optimal barrier \textit{versus} the effective mass and the tunnel barrier offset.

\section*{Acknowledgement}

Nicolas Cavassilas and Daniel Suchet thank the Japan Society for the
Promotion of Science (JSPS) for financial support.

\bibliography{bib_collection}


\end{document}